\documentclass[pra,aps,superscriptaddress,notitlepage,longbibliography,twocolumn,nofootinbib,superscriptaddress]{revtex4-2}
\usepackage{amsmath}
\usepackage{amsfonts}
\usepackage{graphicx}
\usepackage{bm}
\usepackage{bbold}
\usepackage{color}
\usepackage{braket}
\usepackage{comment}
\usepackage{titlesec}

\usepackage[dvipsnames]{xcolor}

\usepackage[colorlinks]{hyperref}

\hypersetup{
    colorlinks=true,
    linkcolor=blue,
    filecolor=magenta,      
    urlcolor=magenta,
    citecolor={blue},
    }

\begin{document}

\abovedisplayskip=6pt
\abovedisplayshortskip=6pt
\belowdisplayskip=6pt
\belowdisplayshortskip=6pt

\title{Gauge invariance of the natural lineshape and dissipative dynamics of a two-level atom}
\author{Chris Gustin}
\affiliation{E.\,L.\,Ginzton Laboratory, Stanford University, Stanford, California 94305, USA}
\email{cgustin@stanford.edu}

\begin{abstract}
The calculation of the natural lineshape of an excited two-level atom (TLA) has long been known to be gauge-dependent, with experiments and classical predictions in better agreement with the lineshape calculated with the dipole gauge (-$\mathbf{r} \cdot \mathbf{E}$ coupling).
We show that by using a Coulomb gauge ($\mathbf{ p} \cdot \mathbf{A}$ coupling) Hamiltonian truncated in a manner consistent with the gauge principle, the correct output spectrum can in fact be obtained. For TLAs undergoing dynamics arising from additional Hamiltonian couplings, we also show that the master equation is gauge-invariant under the same conditions of validity as the Born-Markov approximation, despite different gauges having different spectral densities. These results highlight the importance of using correctly truncated gauge-invariant Hamiltonians in input-output theory for accurate frequency-dependent spectra, even in weak coupling regimes.
\end{abstract}

\maketitle
\section{Introduction}
It has long been observed that the emission spectrum of a two-level atom (TLA) spontaneously emitting into free space can take different forms depending on the gauge used to calculate the output photon observables~\cite{Lamb1952Jan}. The probability of detecting a photon with frequency $\omega$ emitted from an excited TLA with transition resonance $\omega_0$ is generally believed to be~\cite{Milonni1989Oct}
\begin{equation}\label{eq:sdp}
S_{\rm ph}(\omega) = \frac{\Gamma_0}{2\pi}\frac{\omega^3/\omega_0^3}{\Gamma_0^2/4 + (\omega-\omega_0)^2},
\end{equation}
where $\Gamma_0$ is the decay rate of the TLA. This result is derived straightforwardly when the dipole gauge is used. In this gauge (the multipolar gauge with the dipole approximation~\cite{Andrews2018Jan}, also called the length gauge or Poincar{\'e} gauge~\cite{Stokes2021Sep}), the photon-atom interaction is proportional to $-\mathbf{d} \cdot \hat{\mathbf{E}}_{\rm F}$, where $\hat{\mathbf{d}}$ is the transition dipole moment of the TLA and $\hat{\mathbf{E}}_{\rm F}$ is the electric displacement field.
In contrast, in the Coulomb gauge (also called the velocity gauge), where the interaction term is proportional to $ \omega_0\mathbf{d} \cdot \hat{\mathbf{A}}$, straightforward calculations lead to an (incorrect) emission spectrum 
\begin{equation}\label{eq:scp}
S'_{\rm ph}(\omega) = \frac{\Gamma_0}{2\pi}\frac{\omega/\omega_0}{\Gamma_0^2/4 + (\omega-\omega_0)^2}.
\end{equation}
Different gauge calculations leading to predictions differing by squared factors of the ratio of atomic frequencies to electromagnetic ones are commonly encountered when comparing calculations in the dipole gauge vs. the Coulomb gauge, and theoretical approaches to resolve the discrepancy have long been debated~\cite{Lamb1952Jan, Albert1959Sep,Starace1971Apr, Fried1973Dec, Kobe1979Jan,Aharonov1979Oct,Feuchtwang1984Jan,Lamb1987Sep,Milonni1989Oct,Rzazewski2004May,Funai2019Mar}. For example, the induced transition rate of the metastable $2S_{\frac{1}{2}}$ state in hydrogen when weakly coupled with a rf field to the $2P_{\frac{1}{2}}$ state, as in Lamb's famous experiments which measured the eponymous shift~\cite{Lamb1950Aug,Lamb1951Jan}, also sees a similar difference factor in theoretical predictions~\cite{Lamb1987Sep}. Analysis of Lamb's experiments has led to conclusions that the dipole gauge gave better agreement with Lamb's experiment, and should also point to $S_{\rm ph}$ as the correct spectrum~\cite{Lamb1952Jan,Albert1959Sep}. Moreover, simple classical calculations based entirely on electromagnetic fields with no reference to any choice of gauge are in agreement with Eq.~\eqref{eq:sdp}.

In the case of Lamb's experiment, the difference is typically explained in terms of the gauge-dependent definition of atomic-level observables and probability amplitudes as atoms pass through the rf interaction in the experiment~\cite{Funai2019Mar,Lamb1987Sep}. In contrast, the natural lineshape problem of the difference between $S_{\rm ph}$ and $S'_{\rm ph}$ has been explained using varying formalisms, from unphysical precursors present in the Coulomb gauge calculation~\cite{Albert1959Sep}, to analysis of the excitation condition of the TLA~\cite{BibEntry2013Mar,Stokes2013Jun}.

One simple explanation which removes much of the ambiguity in the gauge calculation was provided by Ref.~\cite{Milonni1989Oct}, wherein a proper consideration of the initial condition in which the $\mathbf{A}$ field is ``turned on'' is accounted for. By assuming the TLA to be excited after this (unphysical) turn-on, and including the often neglected $\hat{\mathbf{A}}^2$ term, the authors show that the correct $S_{\rm ph}(\omega)$ spectrum can be obtained from an atomic system. However, their result requires one to truncate the atom to two levels only at the final stage of calculation; working from an initial Hamiltonian which is truncated to a TLA does not give the correct spectrum.

This difficulty recalls recent ``gauge ambiguities'' encountered in cavity quantum electrodynamics (QED), where when the strength of coupling between a TLA and cavity mode becomes sufficiently strong (``ultrastrong coupling'') the usual Hamiltonians used to describe such a system become wildly gauge non-invariant~\cite{Starace1971Apr,Keeling2007Jun,DeBernardis2018Nov}. These problems were solved by Ref.~\cite{DiStefano2019Aug}, (further elaborated on by Refs.~\cite{Savasta2021May,Taylor2020Sep,Taylor2022Mar,Settineri2021Apr,Gustin2023Jan}), in which it was identified that to preserve gauge invariance, the gauge principle must be respected at the level of the introduction of the gauge transformation by ensuring that a local phase
transformation acts on the state vector only at discrete
“lattice” points in space---for a TLA, a two-site lattice. Effectively, this corresponds to introducing interactions between the TLA and field by means of a unitary transformation with an argument truncated to the two-level subspace. This procedure leads to modified (``corrected'') fundamental light-matter Hamiltonians compared with those obtained by projecting the typical minimal coupling Hamiltonians into a two-state subspace, particularly in the Coulomb gauge.

In this work, we show that using the corrected form of the Coulomb Hamiltonian, we are able to derive the correct form of the TLA lineshape $S_{\rm ph}(\omega)$ directly. The corrected Coulomb gauge contains terms proportional to $\hat{\mathbf{A}}$ to all orders, and a self-consistent treatment of these terms leads to the correct Heisenberg equation of motion for the output photon fields, which otherwise are not obtained.
This result highlights the utility of preserving gauge invariance in truncated Hamiltonians beyond the regime of ultrastrong coupling.

The layout of the rest of the paper is as follows: after briefly introducing the formalism of gauge-invariant macroscopic QED in Sec.~\ref{sec:formalism}, we calculate in Sec.~\ref{sec:espectra} the emission spectrum in the dipole gauge, recovering the $S_{\rm ph}(\omega)$ result. We then calculate using the naively truncated Coulomb gauge, which gives $S'_{\rm ph}(\omega)$, followed by our correct calculation using the full Hamiltonian. Following this, we include in Sec.~\ref{sec:master} an analysis of the gauge invariance of the quantum master equation giving rise to dissipation for the TLA when internal or external dynamics are also present. We find that, despite different gauges having different spectral densities, the master equations are gauge invariant in all regimes where the Born-Markov approximation inherent in the master equation is satisfied. We conclude in Sec.~\ref{sec:conclusions}. These results may serve as a useful reference in developing input-output theories in regimes of cavity QED where ultrastrong coupling (or otherwise broadband interactions) between light and matter are present. We also show in Appendix~\ref{app:A} that a simple manifestly gauge-invariant classical calculation leads to a spectrum in agreement with $S_{\rm ph}(\omega)$.
\section{Formalism}\label{sec:formalism}
To calculate the TLA emission spectrum in the Coulomb and dipole gauges, we use the formalism of macroscopic QED, which can describe inhomogeneous linear dielectrics described by dielectric function with real and imaginary parts $\epsilon(\mathbf{r},\omega) = \epsilon_R(\mathbf{r},\omega) + i \epsilon_I(\mathbf{r},\omega)$, encompassing dispersion and absorption. Macroscopic QED can be quantized in a gauge-invariant manner~\cite{Gustin2023Jan}. 
The utility of a formulation in terms of gauge-invariant macroscopic QED allows for easy adaptability of our results to complex photonic environments. We consider only the transverse fields in this paper, which are the ones involved in radiation.

 Under these assumptions, the Schr{\"o}dinger picture fields are
\begin{subequations}
    \begin{equation}
        \hat{\mathbf{E}}_{\rm F}(\mathbf{r}) = i \int \!\! d^3r'  \! \! \int \!\! d\omega \sqrt{\frac{\hbar\epsilon_I(\mathbf{r}',\omega)}{\pi\epsilon_0}}\mathbf{G}(\mathbf{r},\mathbf{r}',\omega) \cdot \hat{\mathbf{b}}(\mathbf{r}',\omega) + \text{H.c.}
    \end{equation}
        \begin{equation}
        \hat{\mathbf{A}}(\mathbf{r}) =  \int \!\! d^3r'  \! \! \int \!\! \frac{d\omega}{\omega} \sqrt{\frac{\hbar\epsilon_I(\mathbf{r}',\omega)}{\pi\epsilon_0}}\mathbf{G}(\mathbf{r},\mathbf{r}',\omega) \cdot \hat{\mathbf{b}}(\mathbf{r}',\omega) + \text{H.c.},
    \end{equation}
\end{subequations}
where $[\hat{b}_i(\mathbf{r},\omega),\hat{b}^{\dagger}_j(\mathbf{r}'\omega')]=\delta_{ij}\delta(\mathbf{r}-\mathbf{r}')\delta(\omega-\omega')$, and $\hat{\mathbf{E}}_{\rm F}$ is the component of the \emph{total} transverse electric field operator $\hat{\mathbf{E}}$ which can be described in terms of bosonic operators; in full, the transverse electric field is $\hat{\mathbf{E}} = \hat{\mathbf{E}}_{\rm F} - \hat{\mathbf{P}}_{\perp}/\epsilon_0$, where $\hat{\mathbf{P}}_{\perp}$ is a transverse, gauge-dependent polarization~\cite{PhysRevA.70.053823,Gustin2023Jan}. The polarization does not contribute to the output spectrum and we neglect it going forwards. The transverse free space photonic Green function $\mathbf{G}$ is well known~\cite{Scheel1999Nov}, and we employ its far-field form later.
Note that even considering free space with a dielectric function $\epsilon(\mathbf{r},\omega)=1$, the fields are expressed in terms of a (fictitious) imaginary part of the dielectric function; formally, one can consider a dielectric function with an imaginary component which vanishes upon taking a limit after all spatial integrals involving the permittivity have been evaluated~\cite{franke2020fluctuation}. No physical predictions depend on this fictitious imaginary component of the permittivity. 

Interactions within the dipole approximation between a TLA with free Hamiltonian $H_{\rm a} = \hbar\omega_0\hat{\sigma}_z/2$ (in terms of the Pauli operators with $\hat{\sigma}^-=\ket{1}\bra{0}$ for ground $\ket{0}$ and $\ket{1}$ excited states) and electromagnetic (EM) fields can be correctly introduced in a gauge-invariant manner by means of the Hamiltonians for the Coulomb and dipole gauges, respectively~\cite{Gustin2023Jan},
\begin{subequations}
\begin{equation}\label{eq:HC_cor}
\hat{H}_{\rm C} = \hat{H}_{\rm F} + \hat{W}\hat{H}_{\rm a}\hat{W}^{\dagger}
\end{equation}
\begin{equation}
\hat{H}_{\rm d} = \hat{W}^{\dagger}\hat{H}_{\rm F}\hat{W} + \hat{H}_{\rm a},
\end{equation}
\end{subequations}
where $\hat{H}_{\rm F} = \int \! d^3 r \! \! \int \!  d\omega \hbar \omega \hat{\mathbf{b}}^{\dagger}(\mathbf{r},\omega) \cdot \hat{\mathbf{b}}(\mathbf{r},\omega)$, and $\hat{W}=\text{exp}\left[i\hat{\Phi}\hat{\sigma}_x/2\right]$, with
\begin{equation}
    \hat{\Phi} = \frac{2}{\hbar} \mathbf{d} \cdot \hat{\mathbf{A}}(\mathbf{0}).
\end{equation}
for a TLA located at the origin with transition dipole moment $\mathbf{d}$ (this formalism is also easily generalized to include multi-level systems and to go beyond the dipole approximation~\cite{Gustin2023Jan}). The two gauges are unitarily equivalent.
\section{Emission spectra of TLA in dipole and Coulomb gauges}\label{sec:espectra}
In this section we calculate the emission spectra (as expressed as a radiative power flow of the EM fields at a far-field surface) of a TLA in free space using different gauge Hamiltonians. First, we use the dipole gauge Hamiltonian which gives the correct spectrum $S_{\rm ph}(\omega)$. Then, we show that the usual (incorrectly truncated) Coulomb gauge Hamiltonian gives the $S'_{\rm ph}(\omega)$ spectra. Finally, using the gauge-corrected Hamiltonian in Eq.~\eqref{eq:HC_cor}, we show that $S_{\rm ph}(\omega)$ is obtained. We use a Heisenberg picture formalism~\cite{PhysRevB.80.195106}, which is easily adapted to input-output theories~\cite{Gardiner1}, and not limited to single-photon subspaces. We neglect any reservoir-induced frequency shifts of the TLA, as these are divergent and beyond the scope of our formalism using non-relativistic and two-level approximations.
\subsection{Dipole gauge}
Considering the dipole gauge Hamiltonian,
\begin{equation}
    \hat{H}_{\rm d} = \hat{H}_{\rm F} + \hat{H}_{\rm a}  - \mathbf{d} \cdot \hat{\mathbf{E}}_{\rm F}(\mathbf{0})\hat{\sigma}_x.
\end{equation}
we calculate the equations of motion of the atom and the field coordinates, respectively, as
\begin{subequations}
\begin{equation}\label{eq:EOM_d_a}
    \dot{\hat{\sigma}}^- = -i\omega_0\hat{\sigma}^- -\frac{i}{\hbar} \mathbf{d} \cdot \hat{\mathbf{E}}_{\rm F}(\mathbf{0})\hat{\sigma}_z
\end{equation}
\begin{equation}\label{eq:EOM_d_b}
    \dot{\hat{\mathbf{b}}}(\mathbf{r},\omega) = -i\omega\hat{\mathbf{b}}(\mathbf{r},\omega)  + \frac{\omega}{2}\mathbf{z}^*(\mathbf{r},\omega)\hat{\sigma}_x,
\end{equation}
\end{subequations}
where 
\begin{equation}
    \mathbf{z}(\mathbf{r},\omega) = \frac{2}{\omega}\sqrt{\frac{\epsilon_I(\mathbf{r},\omega)}{\hbar\pi\epsilon_0}}\mathbf{d}\cdot\mathbf{G}(\mathbf{0},\mathbf{r},\omega).
\end{equation}
We can directly solve Eq.~\eqref{eq:EOM_d_b} in terms of TLA operators,
\begin{equation}\label{eq:EOM_dsolve}
    \hat{\mathbf{b}}(\mathbf{r},\omega,t) = \hat{\mathbf{b}}(\mathbf{r},\omega) e^{-i\omega(t-t_0)}    + \frac{\omega}{2}\mathbf{z}^*(\mathbf{r},\omega)\hat{F}_{\rm d}(\omega,t),
\end{equation}
where we have dropped the counter-rotating $\hat{\sigma}^+(t)$, and defined $\hat{F}_{\rm d}(\omega,t) = \int_{t_0}^{t} dt' \hat{\sigma}^-(t')e^{i\omega(t'-t)}$.
The first term on the right-hand side of Eq.~\eqref{eq:EOM_dsolve} is the ``free'' solution, written in terms of the Schr{\"o}dinger picture operator $\hat{\mathbf{b}}(\mathbf{r},\omega,t_0) = \hat{\mathbf{b}}(\mathbf{r},\omega)$, and does not contribute to any (vacuum) expectation values that will be taken. We thus drop it for the purposes of all following calculations, but it could be included to go beyond the single-photon subspace result.
Under these assumptions, the transverse electric and vector potential fields take the forms:
\begin{subequations}
\begin{equation}
    \hat{\mathbf{E}}_{\rm F}(\mathbf{r},t) = i\int d\omega \frac{\text{Im}\{\mathbf{G}(\mathbf{r},\mathbf{0},\omega)\} \cdot \mathbf{d}}{\pi \epsilon_0} \hat{F}_{\rm d}(\omega,t) + \text{H.c.}
\end{equation}
\begin{equation}
        \hat{\mathbf{A}}(\mathbf{r},t) = \int \frac{d\omega}{\omega} \frac{\text{Im}\{\mathbf{G}(\mathbf{r},\mathbf{0},\omega)\} \cdot \mathbf{d}}{\pi \epsilon_0} \hat{F}_{\rm d}(\omega,t) + \text{H.c.},
\end{equation}
\end{subequations}
where we have used the Green function identity
\begin{equation}
\int d^3r \epsilon_I(\mathbf{r},\omega)\mathbf{G}(\mathbf{r}_1,\mathbf{r},\omega) \cdot \mathbf{G}^*(\mathbf{r},\mathbf{r}_2,\omega) = \text{Im}\{\mathbf{G}(\mathbf{r}_1,\mathbf{r}_2,\omega\}.
\end{equation}
From this, we can calculate the radiative power flow outside of a surface $\mathcal{S}$ enclosing the TLA
\begin{equation}
    P_{\rm rad}(t) = \frac{1}{\mu_0} \oint_{\mathcal{S}} d^2s \hat{s} \cdot \langle :\hat{\mathbf{E}}_{\rm F}(\mathbf{s},t) \times {\bm \nabla} \times \hat{\mathbf{A}}(\mathbf{s},t) :\rangle,
\end{equation}
where we have expressed the power flow in terms of the normally-ordered quantum Poynting vector, to avoid calculating vacuum energy fluctuations.
For simplicity, we choose the surface $\mathcal{S}$ to be a sphere centered around the origin, with a surface in the far-field, such that 
\begin{subequations}
\begin{equation}
    \mathbf{G}(\mathbf{s},\mathbf{0},\omega) = \frac{\omega^2}{4\pi c^2 |\mathbf{s}|}e^{i\omega |\mathbf{s}|/c}\left[\mathbf{1} - \hat{s}\hat{s}\right],
\end{equation}
\begin{equation}
   {\bm \nabla}_{\mathbf{s}} \times  \mathbf{G}(\mathbf{s},\mathbf{0},\omega) = i\frac{\omega}{c} \hat{s} \times  \mathbf{G}(\mathbf{s},\mathbf{0},\omega). 
\end{equation}
\end{subequations}
This gives,
\begin{align}\label{eq:P_rad2}
        P_{\rm rad}(t) &= -i\frac{8\mathbf{d}^2c}{3\pi \epsilon_0} \! \int \! \!  d\omega \! \! \int \! \! d \omega' \!  \beta(\omega) \beta(\omega') \times \nonumber \\ &\sin{\left(\omega |\mathbf{s}|/c\right)}\cos{\left(\omega' |\mathbf{s}|/c\right)} \langle \hat{F}_{\rm d}^{\dagger}(\omega,t) \hat{F}_{\rm d}(\omega',t)\rangle +\text{c.c},
\end{align}
where $\beta(\omega) = \omega^2/(4\pi c^2)$. There are also in principle contribution proportional to $\langle \hat{F}_{\rm d}(\omega,t) \hat{F}_{\rm d}(\omega',t)\rangle$ and $\langle \hat{F}_{\rm d}^{\dagger}(\omega,t) \hat{F}^{\dagger}_{\rm d}(\omega',t)\rangle$, but these are always rapidly oscillating and do not contribute to the radiative power flow in the far-field. The sinusoidal factors in Eq.~\eqref{eq:P_rad2} correspond to retardation effects at retarded time $t_{\rm r} = |\mathbf{s}|/c$, and will, for a far-field surface, oscillate rapidly enough to cause the power flow to vanish, except if compensated for by the exponential factors in $\hat{F}_{\rm d}(\omega,t)$ and $\hat{F}_{\rm d}^{\dagger}(\omega',t)$. Specifically, we can write
\begin{align}\label{eq:Frot}
    &\hat{F}_{\rm d}^{\dagger}(\omega,t)\sin{\left(\omega |\mathbf{s}|/c\right)} \nonumber \\ 
    &=\frac{1}{2i} \int_{t_0}^{t} dt' \hat{\sigma}^+(t')e^{-i\omega(t'-t)}\left[e^{i\omega t_{\rm r}} - e^{-i\omega t_{\rm r}}\right].
\end{align}
The first term in square brackets of Eq.~\eqref{eq:Frot} gives a contribution which is always rapidly oscillating, and vanishes. The second term, in contrast, gives a slowly-varying contribution to the integral when $t' \sim t - t_{\rm r}$. We can also thus extend the upper limit of the integral to $t \rightarrow \infty$, giving
\begin{equation}
\hat{F}_{\rm d}^{\dagger}(\omega,t)\sin{\left(\omega |\mathbf{s}|/c\right)} \approx \frac{i}{2}e^{-i\omega t_{\rm r}}\int_{t_0}^{\infty} dt' \hat{\sigma}^+(t')e^{-i\omega(t'-t)},
\end{equation}
and, similarly,
\begin{equation}
\hat{F}_{\rm d}(\omega',t)\cos{\left(\omega' |\mathbf{s}|/c\right)} \approx \frac{1}{2}e^{i\omega' t_{\rm r}}\int_{t_0}^{\infty} dt' \hat{\sigma}^+(t')e^{i\omega'(t'-t)}.
\end{equation}
We can now consider the time-integrated transmitted power $\int_{t_0}^{\infty} dt P_{\rm rad}(t)$. This gives the time integral $\int_{t_0}^{\infty} dt e^{i(\omega-\omega')(t-t_{\rm r})}$, for which we can extend the lower integral bound to $-\infty$, yielding $2\pi\delta(\omega-\omega')$. Thus,
\begin{equation}
        \int_{t_0}^{\infty} dt P_{\rm rad}(t) = \int d\omega \frac{\hbar \Gamma_0 \omega^4}{2\pi \omega_0^3} S_0(\omega),
\end{equation}
and we have introduced the TLA decay rate $\Gamma_0 = \mathbf{d}^2\omega_0^3/(3\pi \epsilon_0 \hbar c^3)$, and 
\begin{equation}
    S_0(\omega)  = \int_{t_0}^{\infty} dt \int_{t_0}^{\infty} dt' \langle \hat{\sigma}^+(t)\hat{\sigma}^-(t')\rangle e^{i\omega (t'-t)}.
\end{equation}
Finally, assuming the Born-Markov (or equivalently, Wigner-Weisskopf) approximation for the TLA dynamics~\cite{breuer2002theory}, we can use the free-space decay solution $S_0(\omega) = \left[\Gamma_0^2/4 + (\omega-\omega_0)^2\right]^{-1}$, and defining implicitly a photon number spectrum as $\int_{t_0}^{\infty} dt P_{\rm rad}(t) = \int d\omega \frac{S_{\rm ph}(\omega)}{\hbar \omega}$, we find that $S_{\rm ph}(\omega)$ takes the form given in Eq.~\eqref{eq:sdp}. Note $S_{\rm ph}(\omega)$ is only valid in a frequency window around $\omega_0$; making a Markov approximation $\omega^3/\omega_0^3 \approx 1$ under the integral leads to $\int \! d\omega S_{\rm ph}(\omega) =1$ as expected. Note that if the counter-rotating terms in Eq.~\eqref{eq:EOM_d_b} were retained, following the above derivation it can be easily shown that the only change to the emission spectrum is the modification 
 \begin{equation}
     S_0(\omega) \rightarrow S_0(\omega) + \int_{t_0}^{\infty} dt \int_{t_0}^{\infty} dt' \langle \hat{\sigma}^+(t')\hat{\sigma}^-(t)\rangle e^{i\omega (t'-t)},
 \end{equation}
 where the additional term would evaluate under the Born-Markov approximation to $\left[\Gamma_0^2/4 + (\omega+\omega_0)^2\right]^{-1}$. However, in the ultrastrong coupling regime where this term is non-negligible (i.e., when $\Gamma_0$ approaches $\sim \omega_0$) the Born-Markov approximation is no longer valid, and the two-time correlation of the TLA operators would need to be evaluated using other methods (for example, the dynamical polaron ansatz or matrix product states~\cite{PhysRevA.99.013807,PhysRevA.93.043843}). The same considerations apply in the correctly-truncated Coulomb gauge.

\subsection{Naively-truncated Coulomb gauge}
Prior to truncation to a TLA, the Coulomb gauge Hamiltonian is, within the dipole approximation,
\begin{equation}\label{eq:HC_wrong}
\hat{H}_{\rm C,full} = \sum_{\alpha}\frac{\left[\hat{\mathbf{p}}_{k} - q_{k}\hat{\mathbf{A}}(\mathbf{0})\right]^2}{2m_{k}} + \sum_{k,k'}\hat{V}_{\rm Coul}(\hat{\mathbf{r}}_{k},\hat{\mathbf{r}}_{k'}) + \hat{H}_{\rm F},
\end{equation}
where $k$ indexes the particles composing the TLA with mass and charge $m_{k}$ and $q_{k}$, and $\hat{V}_{\rm Coul}$ is the inter-particle Coulomb interaction. Truncating to a two-level system, one calculates the matrix elements of $\hat{\mathbf{p}}$ by using the relation $\hat{\mathbf{p}}_{k} = m_{k}[\hat{\mathbf{r}}_{k},\hat{H}_0]/(i\hbar)$, where $\hat{H}_0 = \sum_{k} \frac{\hat{\mathbf{p}}_{k}^2}{2m_{k}} + \sum_{k,k'}\hat{V}_{\rm Coul}(\hat{\mathbf{r}}_{k},\hat{\mathbf{r}}_{k'})$, and under truncation, $\hat{H}_0 = \hbar \omega_0\hat{\sigma}_z/2$. This gives
\begin{equation}
    \hat{H}'_{\rm C} = \hat{H}_{\rm F} + \frac{\hbar\omega_0}{2}\hat{\sigma}_x  + \omega_0 \mathbf{d} \cdot \hat{\mathbf{A}}(\mathbf{0})\hat{\sigma}_y + \xi_0 \hat{\mathbf{A}}^2(\mathbf{0}),
\end{equation}
where $\xi_0 = \sum_{k}q_{k}^2/(2m_{k})$, and we use a prime on the Hamiltonian to indicate this is the incorrectly-truncated form. This Hamiltonian generates equations of motion:
\begin{subequations}
\begin{equation}\label{eq:EOM_a1w}
    \dot{\hat{\sigma}}^- = -i\omega_0\hat{\sigma}^- + \frac{\omega_0}{\hbar} \mathbf{d} \cdot \hat{\mathbf{A}}(\mathbf{0}) \hat{\sigma}_z
\end{equation}
\begin{align}\label{eq:EOM_a2w}
    \dot{\hat{\mathbf{b}}}(\mathbf{r},\omega) =& -i\omega\hat{\mathbf{b}}(\mathbf{r},\omega)  - i\frac{\omega_0}{2}\mathbf{z}^*(\mathbf{r},\omega) \hat{\sigma}_y \nonumber \\ 
    & - i\frac{\xi_0}{\omega}\sqrt{\frac{\epsilon_I(\mathbf{r},\omega)}{\hbar\pi\epsilon_0}}\hat{\mathbf{A}}(\mathbf{0}) \cdot \mathbf{G}(\mathbf{0},\mathbf{r},\omega).
\end{align}
\end{subequations}
If one neglects the final term in Eq.~\eqref{eq:EOM_a2w}, then, dropping the counter-rotating term, one obtains a solution for the photon operators which is the same as Eq.~\eqref{eq:EOM_dsolve}, but with an additional factor of $\omega_0/\omega$. Propagating this difference through the calculation for the emission spectrum, it easy to see that this leads to the incorrect spectrum $S'_{\rm ph}(\omega)$.

Ref.~\cite{Milonni1989Oct} ``fixes'' this result, by effectively replacing the $\xi_0\hat{\mathbf{A}}^2$ term in Eq.~\eqref{eq:HC_wrong} with $\frac{\omega_0}{\hbar}(\hat{\mathbf{A}}(\mathbf{0}) \cdot \mathbf{d})^2$. However, this result is not only not justified by any physical argument, but also only works if the Heisenberg operators are assumed to act only on a single-excitation subspace: specifically, using the aforementioned second-order replacement term, this leads to a term in the equation of motion for $\dot{\hat{\mathbf{b}}}(\mathbf{r},\omega)$ that is equal to $-i(\omega_0/\hbar)(\mathbf{d}\cdot\hat{\mathbf{A}})\mathbf{z}^*$. Within the one excitation subspace, any matrix element of $\hat{\mathbf{A}}$ can be replaced with $-\hat{\mathbf{A}}\hat{\sigma}_z$; upon this replacement, the approach shown in the following section can then be used to obtain the correct equation of motion.

\subsection{Correctly-truncated Coulomb gauge}
Now, we consider the correctly-truncated Coulomb Hamiltonian for a two-level atom interacting with the transverse EM fields:
\begin{equation}
    \hat{H}_{\rm C} = \hat{H}_{\rm F} + \frac{\hbar \omega_0}{2}\left[ \cos{(\hat{\Phi})}\hat{\sigma}_z + \sin{(\hat{\Phi})}\hat{\sigma}_y\right].
\end{equation}
From here, we calculate the equations of motion of the atom and the field coordinates, respectively, as
\begin{subequations}
\begin{equation}\label{eq:EOMa}
    \dot{\hat{\sigma}}^- = -i\omega_0 \cos{(\hat{\Phi})}\hat{\sigma}^- + \frac{1}{2}\omega_0 \sin{(\hat{\Phi})}\hat{\sigma}_z
\end{equation}
\begin{align}\label{eq:EOMb}
    \dot{\hat{\mathbf{b}}}(\mathbf{r},\omega) = -&i\omega\hat{\mathbf{b}}(\mathbf{r},\omega)  \nonumber \\ & + i\frac{\omega_0}{2}\mathbf{z}^*(\mathbf{r},\omega)\left[\sin{(\hat{\Phi})}\hat{\sigma}_z - \cos{(\hat{\Phi})}\hat{\sigma}_y\right],
\end{align}
\end{subequations}
We can eliminate $\hat{\sigma}_z$ in Eq.~\eqref{eq:EOMb} by using Eq.~\eqref{eq:EOMa} and its conjugate, which gives
\begin{equation}\label{eq:EOM_sub}
        \dot{\hat{\mathbf{b}}}(\mathbf{r},\omega) = -i\omega\hat{\mathbf{b}}(\mathbf{r},\omega)  + i\frac{1}{2}\mathbf{z}^*(\mathbf{r},\omega)\left(\dot{\hat{\sigma}}^- + \dot{\hat{\sigma}}^+\right).
\end{equation}
Dropping the counter-rotating term, we obtain,
\begin{equation}\label{eq:EOM_bsolve_C}
    \hat{\mathbf{b}}(\mathbf{r},\omega,t) = \hat{\mathbf{b}}(\mathbf{r},\omega) e^{-i\omega(t-t_0)}    + \frac{i}{2}\mathbf{z}^*(\mathbf{r},\omega)\hat{F}_{\rm C}(\omega,t).
\end{equation}
We again drop the free vacuum term going forward. We have also defined:
\begin{align}
    &\hat{F}_{\rm C}(\omega,t) = \int_{t_0}^t dt' \dot{\hat{\sigma}}^-(t') e^{i\omega(t'-t)} \nonumber \\ 
    & = \hat{\sigma}^-(t) - \hat{\sigma}^-(t_0)e^{i\omega(t_0-t)} - i\omega\int_{t_0}^t dt' \hat{\sigma}^-(t')e^{i\omega(t'-t)} \nonumber \\
    & =\hat{\sigma}^-(t) - i\omega\int_{t_0}^t dt' \hat{\sigma}^-(t')e^{i\omega(t'-t)},
\end{align}
where, in the third line, we have let $\hat{\sigma}^-(t_0) = 0$, corresponding to an excitation of the TLA at some time $t_1 > t_0$ after the atom-field interaction is ``switched-on''. Without this assumption, the unphysical ``switching-on'' of the interaction can lead to an incorrect form of the emission spectrum~\cite{Milonni1989Oct}.
We note that one can also expand the Hamiltonian $\hat{H}_{\rm C}$ to a given order in $\mathbf{d} \cdot \hat{\mathbf{A}}$ (but at least $\mathcal{O}(\mathbf{d}^2)$), and perform the calculation that way. 

Under these assumptions, we calculate the radiative power flow through a far-field surface
\begin{align}\label{eq:P_a}
        P_{\rm rad}(t) &= -i\frac{8\mathbf{d}^2c}{3\pi \epsilon_0} \! \int \! \!  \frac{d\omega}{\omega} \! \! \int \! \! \frac{d \omega'}{\omega'} \!  \beta(\omega) \beta(\omega') \times \nonumber \\ &\sin{\left(\omega |\mathbf{s}|/c\right)}\cos{\left(\omega' |\mathbf{s}|/c\right)} \langle \hat{F}_{\rm C}^{\dagger}(\omega,t) \hat{F}_{\rm C}(\omega',t)\rangle +\text{c.c}.
\end{align}
Equation~\eqref{eq:P_a} is the same as the dipole gauge result in Eq.~\eqref{eq:P_rad2}, but with additional factors of $\frac{1}{\omega}$ and $\frac{1}{\omega'}$, and $\hat{F}^{(\dagger)}_{\rm C}$ instead of $\hat{F}^{(\dagger)}_{\rm d}$. Crucially, the $\hat{\sigma}^-(t)$ term in $\hat{F}_{\rm C}(\omega',t)$ contains no term oscillating at frequency $\omega'$ to cancel out the effect of retardation, and thus, within the radiative power flow, we can take $\hat{F}_{\rm C}(\omega',t) \rightarrow -i\omega' \hat{F}_{\rm d}(\omega',t)$, and similarly for its conjugate $\hat{F}^{\dagger}_{\rm C}(\omega,t) \rightarrow i \omega \hat{F}_{\rm d}(\omega,t)$, recovering the result of $S_{\rm ph}(\omega)$. We note here that to calculate the atomic dynamics, we again assume the Born-Markov approximation, and here it is sufficient to expand the equations of motion for the TLA subspace to leading order in $\mathbf{d} \cdot \hat{\mathbf{A}}$, consistent with the Born-Markov approximation and neglecting (divergent) frequency shifts.

\section{Gauge invariance of master equation under internal atomic dynamics}\label{sec:master}
Next, we show that gauge invariance is preserved even under interactions. This is a non-trivial point, as the decay rate $\Gamma_0$ which we have used in the previous section no longer applies in the presence of interactions. Properly, one must go into the interaction picture and calculate the new atomic level eigenstates to compute the new dissipator in the presence of interactions. However, the spectral densities are different in the two gauges (in the Coulomb gauge, the spectral density is reduced by a factor of $\omega^2_0/\omega^2$).
Unlike the previous examples, we express our results in terms of a generic medium Green function, allowing the results to be used for arbitrary photonic structures, including dispersive and absorptive ones.
\subsection{Dipole gauge}
As an example, consider the system with dipole gauge Hamiltonian
\begin{equation}
    \hat{H}_{\rm d} = \hat{H}_0 + \hat{V} - \mathbf{d} \cdot \hat{\mathbf{E}}_{\rm F}(\mathbf{0}) \hat{\sigma}_x,
\end{equation}
where
\begin{equation}
\hat{H}_0 = \hat{H}_{\rm F} + \hbar \omega_0 \hat{\sigma}^+ \hat{\sigma}^- + \sum_j \hbar \omega_j \hat{A}_j^{\dagger}\hat{A}_j
\end{equation}
is a diagonal Hamiltonian,
 where $\hat{A}^{\dagger}$ and $\hat{A}$ correspond to excitation operators indexed by $j$ with excitation energy $\omega_j$; we keep this generic, but these could correspond to different levels of the same atom, bosonic oscillators modes, or different particles, for examples. The Hamiltonian $\hat{V}$ couples the system of the TLA with the other oscillator states. For simplicity, only the TLA couples to the radiation field. For other interactions where $\hat{V}$ has transverse EM origin, a similar formalism could be used, but care would need to be taken to ensure that the gauge transformation operator $\hat{\Phi}$ also incorporates this coupling to the radiation field. We stress that it is assumed here that $\hat{V}$ is already a correct representation of the TLA interactions in the dipole gauge once truncating to two atomic levels; for a more rigorous approach, one could start from a fundamental Lagrangian including this interaction and then introduce gauge-invariant light-matter coupling, as in Ref.~\cite{Gustin2023Jan} for example.
It is easy to verify for this Hamiltonian that the Heisenberg solution in Eq.~\eqref{eq:EOM_dsolve} also applies to this system.

Now, we can calculate the master equation for this system, as $\dot{\hat{\rho}} = -\frac{i}{\hbar}[\hat{H}_{\rm S},\hat{\rho}] + \mathcal{L}\hat{\rho}$, where the dissipator is 
\begin{equation}
\mathcal{L}\hat{\rho}=\frac{1}{{2\pi}}\int_0^{\infty}\! \!\! d\tau  \! \int \! d\omega \Gamma(\omega)e^{-i\omega \tau}[\hat{\sigma}_x(-\tau)\hat{\rho},\hat{\sigma}_x] + \text{H.c.}
\end{equation}
where $\Gamma_0$ is the bare decay rate, and $\sigma^-(-\tau) = e^{-\frac{i}{\hbar} \hat{H}_{\rm S}\tau}\hat{\sigma}^-e^{\frac{i}{\hbar}\hat{H}_{\rm S}\tau}$, where $\hat{H}_{\rm S} = \hat{H}_0 + \hat{V}$ is the system Hamiltonian. The (dipole gauge) spectral density here is proportional to 
\begin{equation}
\Gamma(\omega) = \frac{2}{\hbar\epsilon_0} \mathbf{d} \cdot \text{Im}\{\mathbf{G}(\mathbf{0},\mathbf{0},\omega)\} \cdot \mathbf{d}.
\end{equation}
Moving to the basis of system Hamiltonian eigenstates, one obtains (again neglecting frequency shifts)
\begin{equation}\label{eq:diss_d}
\mathcal{L}\hat{\rho} = \frac{1}{2}\sum_{\alpha \alpha'}\Gamma(\omega_{\alpha})c_{\alpha}c^*_{\alpha'}[\hat{\sigma}_{\alpha}\hat{\rho},\hat{\sigma}^{\dagger}_{\alpha'}] + \text{H.c.}
\end{equation}
Here, we have use notation in which $\alpha$ is an index that refers to the pair of eigenstates of $\hat{H}_{\rm S}$ with indices $(j,k)$, such that the energy of the state $k$ is greater than that of the state $j$, and $c_{\alpha} = \bra{j}\hat{\sigma}_{x}\ket{k}$,  $\hat{\sigma}_{\alpha} = \ket{j}\bra{k}$, and $\omega_{\alpha} = \omega_k - \omega_j$.
\subsection{Coulomb gauge}\label{sec:coulomb_me}
Now consider the correctly-truncated Coulomb gauge Hamiltonian $\hat{H}_{\rm C} = \hat{W} \hat{H}_{\rm d} \hat{W}^{\dagger}$. Without any approximations, the Hamiltonian is
\begin{equation}
    \hat{H}_{\rm C} = \hat{H}_{\rm F} + \frac{\hbar \omega_0}{2}\left[ \cos{(\hat{\Phi})}\hat{\sigma}_z + \sin{(\hat{\Phi})}\hat{\sigma}_y\right] + \hat{V}',
\end{equation}
where 
\begin{equation}
\hat{V}' = \text{exp}\left[i\hat{\Phi}\hat{\sigma}_x/2\right]\hat{V}\text{exp}\left[-i\hat{\Phi}\hat{\sigma}_x/2\right].
\end{equation}
It is straightforward to show that the solution for the Heisenberg operator $\hat{\mathbf{b}}(\mathbf{r},\omega,t)$ from Eq.~\eqref{eq:EOM_bsolve_C} is unchanged by the addition of the $\hat{V}'$ term.

Next, we expand the fields to leading order in $\mathbf{d} \cdot \hat{\mathbf{A}}(\mathbf{0})$, consistent with the Born-Markov approximation and neglecting energy shifts. This gives
\begin{align}\label{eq:C_w_V}
    \hat{H}_{\rm C} =  \hat{H}_0 + \omega_0 \mathbf{d} \cdot \hat{\mathbf{A}}(\mathbf{0}) \hat{\sigma}_y + \hat{V} + \frac{i}{\hbar}\mathbf{d} \cdot \hat{\mathbf{A}}(\mathbf{0})[\hat{\sigma}_x,\hat{V}].
\end{align}

The Coulomb gauge has the same system Hamiltonian $\hat{H}_{\rm S}$, and so we can calculate the Coulomb gauge master equation as
\begin{equation}
\mathcal{L}'\hat{\rho}=\frac{1}{{2\pi}}\int_0^{\infty}\! \!\! d\tau  \! \int \! d\omega \Gamma(\omega)\left(\frac{\omega_0^2}{\omega^2}\right)e^{-i\omega \tau}[\hat{S}(-\tau)\rho,\hat{S}] + \text{H.c.},
\end{equation}
where we have used the fact that the spectral density in the Coulomb gauge, coming from the $\mathbf{d} \cdot \hat{\mathbf{A}}$ coupling, is identical to the dipole gauge spectral density except for a factor of $\omega_0^2/\omega^2$, and 
\begin{equation}
\hat{S} = \hat{\sigma}_y + \frac{i}{\hbar \omega_0}[\hat{\sigma}_x,\hat{V}]
\end{equation}
is the system operator which couples to the reservoir. Proceeding, one obtains
\begin{equation}
\mathcal{L}'\hat{\rho} = \frac{1}{2}\sum_{\alpha \alpha'}\Gamma(\omega_{\alpha})\left(\frac{\omega_{0}^2}{\omega_{\alpha}^2}\right)c'_{\alpha}c'^*_{\alpha'}[\hat{\sigma}_{\alpha}\hat{\rho},\hat{\sigma}^{\dagger}_{\alpha'}] + \text{H.c.}
\end{equation}
Next, we note that 
\begin{align}\label{eq:calg}
    c'_{\alpha} &= \bra{j}\left[\hat{\sigma}_y + \frac{i}{\hbar\omega_0}\left[\hat{\sigma}_x,\hat{V}\right]\right]\ket{k} \nonumber \\ 
    & = \bra{j}\left[\hat{\sigma}_y + \frac{i}{\hbar\omega_0}\left[\hat{\sigma}_x,\hat{H}_{\rm S} - \hbar \omega_0 \hat{\sigma}^+\hat{\sigma}^- \right]\right]\ket{k} \nonumber \\ 
    & = \bra{j}\left[  \frac{i}{\hbar\omega_0}\left[\hat{\sigma}_x,\hat{H}_{\rm S} \right]\right]\ket{k} \nonumber \\ 
    & = i\left(\frac{\omega_\alpha}{\omega_0}\right)c_{\alpha}
\end{align}
where in the second line we used that $\hat{A}_j$ and $\hat{A}_{j}^{\dagger}$ excitations commute with the TLA operators. Thus, the dissipator becomes
\begin{equation}\label{eq:diss_c}
\mathcal{L}'\hat{\rho} = \frac{1}{2}\sum_{\alpha \alpha'}\Gamma(\omega_{\alpha})\left(\frac{\omega_{\alpha'}}{\omega_{\alpha}}\right)c_{\alpha}c^*_{\alpha'}[\hat{\sigma}_{\alpha}\hat{\rho},\hat{\sigma}^{\dagger}_{\alpha'}] + \text{H.c.},
\end{equation}
which is almost identical to the dipole gauge result, aside from a factor of $\omega_{\alpha'}/\omega_{\alpha}$. Under the secular approximation, valid when the differences between transitions $|\omega_{\alpha} - \omega_{\alpha'}|$ is much greater than the dissipative rates $\Gamma(\omega_{\alpha})$ for all $\alpha$, the sum is reduced to only terms with $\alpha = \alpha'$. Thus, we conclude that the master equation under the Born-Markov approximation gives gauge-invariant results for a TLA with coupling dynamics when the secular approximation can be performed. This result was also found by Stokes and Nazir~\cite{Stokes2018Apr}, considering the specific case of two dipoles interacting in free-space (which can also be treated in our approach, provided the field coupling to the other dipole is also accounted for in the Hamiltonian and gauge transformation).

We can attribute the failure of gauge invariance to a failure of the conditions of the Born-Markov condition to be satisfied. For cases where $\Gamma(\omega)$ changes on frequency scales which vary more rapidly than powers of $\omega$ (e.g., structured photonic media with resonances), then $\omega_{\alpha}/\omega_{\alpha'}$ will be essentially unity in all nonvanishing regimes of the spectral density, and the master equation is effectively gauge-invariant. On the other hand, if $\Gamma(\omega)$ varies on the scale of $\omega_0$ (i.e., a power of $\omega$, as in an homogeneous medium), as an example consider two transitions $\omega_{\alpha} = \omega_0$ and $\omega_{\alpha'} = \omega_0 + \Delta$. Then, $\omega_{\alpha}/\omega_{\alpha'}$ differs from unity when $\Delta/\omega_0$ is appreciable. If the secular approximation is not satisfied, then $\Delta \sim \Gamma_0$. But if $\Gamma_0/\omega_0 \ll 1$ is not satisfied, then the Born-Markov approximation fails. \emph{Thus, in regimes where the Born-Markov approximation holds, the master equation is gauge-invariant.} Sec.~\ref{sec:ex} illustrates this with an example.

\subsection{Naively truncated Coulomb gauge}
We now consider the gauge invariance (or noninvariance) of the master equation under internal TLA dynamics if the naively truncated Coulomb gauge is instead used. To be explicit, we assume that the previously employed truncated dipole gauge coupling Hamiltonian $\hat{V}$ can be written as
\begin{equation}\label{eq:Vtrunc}
\hat{V} = \mathcal{P}\hat{\mathcal{V}}(\hat{\mathbf{r}},\hat{\mathbf{p}})\mathcal{P},
\end{equation}
where $\mathcal{P} = \sum_{i=0}^1\ket{i}\bra{i}$ is the operator which projects the Hamiltonian to the two-level subspace, and again we assume only the TLA subspace couples to the radiation field, so we can omit dependencies on other operators (the $\hat{A}_j$, $\hat{A}_j^{\dagger}$). Equation~\eqref{eq:Vtrunc} is thus completely general. The untruncated Coulomb gauge Hamiltonian is then, from Eq.~\eqref{eq:HC_wrong}, $\hat{H}_{\rm C, full} \rightarrow \hat{H}_{\rm C,full} + \hat{\mathcal{V}}_{\rm C,full}$, where
\begin{equation}
\hat{\mathcal{V}}_{\rm C,full} = \hat{\mathcal{W}}\hat{\mathcal{V}}(\hat{\mathbf{r}},\hat{\mathbf{p}})\hat{\mathcal{W}}^{\dagger},
\end{equation}
where
\begin{equation}
\hat{\mathcal{W}} = \text{exp}\left[i\frac{q}{\hbar}\hat{\mathbf{r}} \cdot \hat{\mathbf{A}}(\mathbf{0})\right]
\end{equation}
is the untruncated unitary gauge transformation operator (note under truncation, $\mathcal{P}q\hat{\mathbf{r}}\mathcal{P} = \mathbf{d}\hat{\sigma}_x$). Expanding this Hamiltonian to first order in the vector potential and truncating, we obtain
\begin{equation}
\hat{H}_{\rm C}' = \hat{H}_0 + \omega_0 \mathbf{d} \cdot \hat{\mathbf{A}}(\mathbf{0})\hat{\sigma}_y + \hat{V} + \mathcal{P}\left[i\frac{q}{\hbar}\hat{\mathbf{r}} \! \cdot \! \hat{\mathbf{A}}(\mathbf{0}),\hat{\mathcal{V}}(\hat{\mathbf{r}},\hat{\mathbf{p}})\right]\mathcal{P},
\end{equation}
which is identical to Eq.~\eqref{eq:C_w_V}, except for the last term. In general, it is now well known that truncating after the gauge transformation in the untruncated space will lead to different results~\cite{Keeling2007Jun,DeBernardis2018Nov,DiStefano2019Aug,Taylor2020Sep}. As such, the operator algebra that is employed in Eq.~\eqref{eq:calg} will no longer hold. If the coupling $\hat{V}$ is weak, the difference between $c_{\alpha}$ and $c'_{\alpha}$ calculated with the naively truncated Coulomb gauge may be small enough as to maintain gauge invariance, but under sufficiently strong interactions $\hat{V}$ (for example, those approaching ultrastrong, such that $\hat{V} \sim \omega_0$), the differences between the naively truncated Coulomb gauge Hamiltonian and the dipole gauge can generally be expected to lead to a breakdown of gauge invariance, as it is not guaranteed that a secular approximation could be made to remove the influence of the terms causing such a gauge dependence.

\subsection{Coupling to an auxiliary TLA: A simple illustration of gauge invariance under internal atomic dynamics}\label{sec:ex}

As a very simple example to illustrate the arguments about gauge invariance in Sec.~\ref{sec:coulomb_me}, we consider the case where the TLA is coupled to an auxiliary TLA with degenerate transition energy $\hbar \omega_0$. Explicitly, we have
\begin{subequations}
    \begin{equation}
        \hat{H}_{0} = \hat{H}_{\rm F} + \hbar\omega_0 \hat{\sigma}^+\hat{\sigma}^- + \hbar{\omega}_0\hat{\sigma}^+_{\rm a}\hat{\sigma}^-_{\rm a}
    \end{equation}
    \begin{equation}
        \hat{V} = \hbar \Omega \left[\hat{\sigma}^-\hat{\sigma}^+_{\rm a} + \hat{\sigma}^+\hat{\sigma}^-_{\rm a}\right],
    \end{equation}
\end{subequations}
where $\hat{\sigma}^{\pm}_{\rm a}$ are raising and lowering operators of the auxiliary TLA. 

For the dipole gauge and Coulomb gauge dissipators in Eqs.~\eqref{eq:diss_d} and~\eqref{eq:diss_c} to be approximately equivalent, and thus gauge invariance to be retained, we require either that the difference between the two is small, or that the secular approximation can be employed. To quantify this, we propose the transition matrix metric
\begin{equation}
    B_{\alpha\alpha'} = \frac{\Gamma(\omega_{\alpha})\left(\omega_{\alpha'}/\omega_{\alpha} - 1\right)}{\Gamma(\omega_0)}c_{\alpha}c^*_{\alpha'},
\end{equation}
where the denominator is the rate evaluated for the $\hat{V}=0$ case, where $c_{\alpha}=1$.  $|B_{\alpha,\alpha'}| \ll 1$ is a sufficient condition for the master equation to be gauge invariant, as in this instance the difference in magnitude of the rates corresponding to the different gauge master equations is small (relative to the uncoupled case, which we assume does not give a substantially smaller contribution---if it were, a different normalization would need to be used).

The Hamiltonian $\hat{H}_0$ is easily diagonalized, with eigenstates $\ket{\psi_1} = \ket{0}\ket{0}_{\rm a}$, $\ket{\psi_2} = \frac{1}{\sqrt{2}}\left[\ket{0}\ket{1}_{\rm a} - \ket{1}\ket{0}_{\rm a}\right]$, $\ket{\psi_3} = \frac{1}{\sqrt{2}}\left[\ket{0}\ket{1}_{\rm a} + \ket{1}\ket{0}_{\rm a}\right]$, and $\ket{\psi_4} = \ket{1}\ket{1}_{\rm a}$, with eigenenergies $\hbar\omega_1=0$, $\hbar\omega_2 = \hbar\omega_0 - \hbar \Omega$, $\hbar\omega_3 = \hbar\omega_0 + \hbar \Omega$, and $\hbar\omega_4 = 2\hbar\omega_0$. We assume without loss of generality that $\Omega >0$, and require $\Omega < \omega_0$ to ensure the eigenenergies are ordered from least to greatest.

As a first example, we consider the case where the spectral density varies rapidly in frequency. A canonical example of such can be obtained by letting the dipole gauge spectral density correspond to a TLA coupling to a single cavity resonance, such that
\begin{equation}
\Gamma_{\rm cav}(\omega) =  \frac{\omega\omega_0\kappa\eta^2}{(\omega-\omega_0)^2 + \kappa^2/4},
\end{equation}
where we assume for simplicity the cavity frequency to be resonant with the TLA transition $\omega_0$, and $\kappa$ is the cavity linewidth.
Technically, this form of the spectral density arises from coupling to a single lossy cavity mode (quasinormal mode) that is strictly real at the position of the dipole~\cite{Gustin2025,gustin2025dissipationbroadbandultrastrongcoupling}, and $\eta$ is a dimensionless parameter characterizing the strength of the light-matter interaction. To ensure the spectral density varies rapidly in frequency, we assume a high quality factor $Q = \omega_0/\kappa \gg 1$, and for the Born-Markov approximation to hold, we require $\eta Q \ll 1$.  In this basis, we find the only nonzero elements of $B_{\alpha\alpha'}$ to be expressible in the basis $\alpha_1 = (2,1)$, $\alpha_2 = (3,1)$, $\alpha_3=(4,3)$, $\alpha_4 = (4,2)$, such that
\begin{equation}
    B^{\rm cav} = 
	\begin{bmatrix}
0 & -B_0 & 0 & -B_0 \\
B_0 & 0 & -B_0 & 0 \\ 
0 & B_0 & 0 & B_0 \\ 
B_0 & 0 & -B_0 & 0,
\end{bmatrix}
\end{equation}
where 
\begin{equation}
    B_0 = \frac{\tilde{\Omega}}{1+4Q^2\tilde{\Omega}^2},
\end{equation}
with $\tilde{\Omega} = \Omega/\omega_0$. As a function of $\tilde{\Omega}$, $B_0$ takes on a maximum value of $B_{0,\text{max}} = 1/(4Q)$ at $\tilde{\Omega} = 1/(2Q)$. Since $Q \gg 1$, we thus see that the master equation is always gauge invariant. If the cavity $Q$ factor were not large, the spectral density would be effectively slowly varying over the relevant dynamical bandwidth, and the proper analysis of gauge invariance would be more similar to the following example.

As a second example, we can consider a free-space (slowly varying) dipole gauge spectral density, where $\Gamma_{\rm free}(\omega) = \Gamma_0\left(\omega/\omega_0\right)^3$. In this case, the transition matrix metric becomes
\begin{equation}
    B^{\rm free} = 
	\begin{bmatrix}
0 & -B_- & 0 & -B_- \\
B_+ & 0 & -B_+ & 0 \\ 
0 & B_- & 0 & B_- \\ 
B_+ & 0 & -B_+ & 0,
\end{bmatrix}
\end{equation}
where 
\begin{equation}
    B_{\pm} =  \tilde{\Omega}\left(1\pm \tilde{\Omega}\right)^2.
\end{equation}
The criterion for the Born-Markov approximation to hold here is simply that $\Gamma_0 \ll \omega_0$. For the master equation to be gauge invariant, we must either have $|B_{\pm}| \ll 1$, or be in a regime where the secular approximation holds, which in this case is simply $ \Omega \gg \Gamma_0$. Depending on the magnitude of $\Omega$, are three possible scenarios: (i) $\Omega \ll \Gamma_0$, (ii) $\Omega \sim \Gamma_0$, or (iii) $\Omega \gg \Gamma_0$. In case (i), $\tilde{\Omega} = \left(\frac{\Omega}{\Gamma_0}\right)\left(\frac{\Gamma_0}{\omega_0}\right)$, which is the product of two small parameters, and thus clearly $|B_{\pm}| \ll 1$. In case (ii), $\tilde{\Omega}\sim \frac{\Gamma_0}{\omega_0} \ll 1$, and so $|B_{\pm}| \ll 1$ also. Finally, in case (iii), the criterion for the secular approximation is satisfied. Therefore, for any interaction strength $\Omega$, the master equation remains gauge invariant.
\section{Conclusions}\label{sec:conclusions}
In conclusion, we have shown that the correct form of the natural lineshape of a TLA $S_{\rm ph}(\omega)$ can be obtained in the Coulomb gauge, in contrast to previous calculations, \emph{if} the corrected form of the Coulomb gauge Hamiltonian is used, which preserves gauge invariance. As a consequence, correct application of input-output theory in cases where detailed frequency dependence of output observables is desired may also require the use of this corrected Hamiltonian. This is particularly important in highly broadband regimes of quantum optics and cavity QED~\cite{Gustin2025,gustin2025dissipationbroadbandultrastrongcoupling}. We have further shown that master equations used to calculate TLA observables are gauge-invariant, even though the spectral densities corresponding to the TLA-reservoir interaction are gauge-dependent. We expect this work to be a useful aid to formulations of master equations and input-output theory in broadband regimes of coupling between light and matter, in particular with regards to ultrastrong coupling. 

Finally, we remark that the dipole gauge is, in some cases, easier to connect to experimental observables involving the TLA degrees of freedom, as the canonical momentum within the dipole approximation remains equal to the physical momentum even in the presence of interactions. In the Coulomb gauge, more care needs to be taken with regards to initial conditions. However, we note that when fields are adiabatically introduced into the dynamics, either gauge is naturally suited to simply calculating TLA observables~\cite{Milonni1989Oct,Lamb1987Sep}.

\appendix
\section{Classical Calculation of Lineshape}\label{app:A}
In this appendix we present a simple fully classical mechanical derivation of Eq.~\eqref{eq:sdp} for the TLA lineshape $S_{\rm ph}(\omega)$. In this, we shall assume the TLA to be represented by a simple classical oscillator; such an assumption of course neglects quantum mechanical effects, but in the single-quanta regime studied in most of this work no unique quantum mechanical effects have an influence on the emitted lineshape.

For simplicity, we employ a simple definition~\cite{Hughes2024Jun} of the spectrum of a classical scattered fields $S_{\rm cl}(\mathbf{r},\omega) = |\mathbf{E}(\mathbf{r},\omega)|^2$, where $\mathbf{E}(\mathbf{r},\omega)$ is the electric field. Assuming a simple frequency-independent excitation $\mathbf{E}_0$, the scattered field is $\mathbf{E}(\mathbf{r},\omega) = \mathbf{G}(\mathbf{r},\mathbf{0},\omega) \cdot {\bm \alpha}(\omega) \cdot \mathbf{E}_0$, where 
\begin{equation}
{\bm \alpha}(\omega) = \left[1-{\bm \alpha}_0(\omega) \cdot \mathbf{G}(\mathbf{0},\mathbf{0},\omega)\right]^{-1} \cdot {\bm \alpha}_0(\omega),
\end{equation}
is the polarizability of a point scatterer representing the TLA, and
\begin{equation}
   {\bm \alpha}_0(\omega) =  \frac{2 \mathbf{d} \mathbf{d} \omega_0}{\hbar \epsilon_0(\omega_0^2-\omega^2)}
\end{equation}
is the bare polarizability of the TLA with resonance $\omega_0$~\cite{VanVlack2012Apr}. Employing the free space form of the Green's function, we can straightforwardly calculate $S_{\rm cl}(\mathbf{r},\omega)$. The real part of the Green function with both spatial arguments evaluated at $\mathbf{r}=\mathbf{0}$ has a divergence corresponding to the self-energy of a point scatterer~\cite{RevModPhys.70.447}, which we neglect, and its imaginary part is $\text{Im}\{\mathbf{G}(\mathbf{0},\mathbf{0},\omega)\} =\frac{ \omega^3}{6\pi c^3}\mathbf{I}$. Thus, we find (employing the far-field form of the Green function for $\mathbf{G}(\mathbf{r},\mathbf{0},\omega)$)
\begin{equation}
S_{\rm cl}(\mathbf{r},\omega) = F(\mathbf{r}) \omega^4 \Bigg|\frac{1}{\omega_0^2-\omega^2 - i \Gamma_0^2\omega^3/\omega_0^2} \Bigg|^2,
\end{equation}
where $F(\mathbf{r})$ is a frequency-independent factor. As the excitation condition here is different than the quantum model in the main text, the overall magnitude of the spectrum will differ, we can neglect this factor. Additionally, consistent with the rotating-wave approximation made in the main text, we can let $\omega_0^2-\omega^2 \approx 2\omega_0(\omega_0-\omega)$ and $\omega^3/\omega_0^2 \approx \omega_0$ in the denominator for frequencies near $\omega \approx \omega_0$. Thus, dividing by $\omega$ to move from a energy flux spectrum to a photon number spectrum, we find 
\begin{equation}
    \frac{S_{\rm cl}(\mathbf{r},\omega)}{\omega} \propto S_{\rm ph}(\omega).
\end{equation}
This result is clearly gauge-invariant as no potential fields have been introduced.

\acknowledgments
This work was supported by NSF awards PHY-2409353 and CCF-1918549. We thank Stephen Hughes for useful conversations.

\bibliography{main}
\end{document}